\documentclass[%
twocolumn,
 amsmath,amssymb,
 aps, physrev,
]{revtex4-2}
\usepackage[utf8]{inputenc}
\usepackage{babel}
\usepackage{color}
\usepackage{float}
\usepackage{amsbsy}
\usepackage{amstext}
\usepackage{amsmath}
\usepackage{graphicx}
\usepackage{esint}
\usepackage{setspace} 

\begin{document}

\title{\textbf{Pure many-body interactions in colloidal systems by artificial random light fields} 
}%

\author{Augustin Muster}
\author{Luis S. Froufe-Pérez}
 \email{Corresponding author: luis.froufe@unifr.ch}
\affiliation{%
Department of Physics, University of Fribourg, Chemin du Musée 3, 1700 Fribourg, Switzerland
}%

\date{\today}

\begin{abstract}
We propose a method to generate pure many-body interactions in colloidal systems by using optical forces induced by random optical fields with an optimized spectral energy density. To assess the feasibility in general settings, we develop a simple model for Lorentzian  electric and magnetic dipole response. An optimization procedure is then introduced to design the spectral energy density of the random field that minimizes pair interactions at constant electromagnetic energy density. We conclude that, under rather general circumstances, it is possible to effectively cancel pair interactions within a range of distance. Hence a colloid can be driven to interact exclusively through many-body interactions.
\end{abstract}

\maketitle


Understanding and tuning interparticle interactions in colloidal systems is essential for controlling their properties, including dynamics and  assembly capabilities \cite{hunter1987foundations,cosgrove2010colloid}. These interactions, whether of electrostatic origin, such as the electrical double layer \cite{jones2002soft,debye1923theorie}, or of electromagnetic fluctuations origin \cite{israelachvili2011intermolecular,buhmann_dispersion_2013,buhmann2013dispersion}, such as Van der Waals \cite{parsegian2005van} Casimir \cite{casimir1948attraction}, or Casimir–Polder \cite{casimir1948influence} forces, are most often treated as pairwise. A prominent example is the Derjaguin–Landau–Verwey–Overbeek (DLVO) \cite{israelachvili2011intermolecular} theory, which describes the balance of dispersion forces and electrostatic double-layer repulsion as an effective pair potential. However, restricting the description to pairwise interactions is insufficient to capture the full behavior of colloidal systems because many-body effects can play a important role in it \cite{dobnikar2002many,dobnikar2003many,dobnikar2004three,brunner2004direct}.

While pairwise interactions can be tuned through well established techniques such as electrostatic screening or steric stabilization \cite{israelachvili2011intermolecular,derjaguin1940repulsive}, light offers an additional means of control, either via optical tweezers and lattices \cite{Delgado-Buscalioni2018Emergence} , broad band coherent fields \cite{Holzmann2016Tailored}, or random isotropic fields \cite{Marques2016Crossover, Douglass2012Superdiffusion}. In particular, it has been demonstrated that isotropic optical binding (OB) can be induced using artificially generated random optical fields \cite{boyer1973retarded,brugger_controlling_2015,luis-hita_active_2022}, which can be tailored to mimic a desired potential, for instance such as inverse square potential or "mock gravity" \cite{luis-hita_light_2019}. Even coherently scattered black body radiation induce weak attractive forces between atoms and macroscopic objects \cite{Haslinger2017Attractive}. In many of the studied cases, it is coherent scattering that allows for the necessary interferences that build the characteristics of the force as a function of the distance between scatterers.
However, studying and controlling many-body interactions is far more challenging. Understanding them requires disentangling their contributions from the pair interactions. Ideally, suppressing pair interactions altogether would give access to systems interacting exclusively through many-body interactions.

In this letter, we investigate the possibility of developing this strategy by exploiting optical interactions induced by random fields in colloidal systems. Since these interactions originate from multiple scattering, they are intrinsically of a many-body nature, whereas it is well established \cite{brugger_controlling_2015} that pairwise contributions can be tuned, to some extent, by adjusting the spectral energy density $u_E(\omega)$ of the artificial random field. In order to include the necessary interference of the coherently scattered fields, we model colloidal particles as electric and magnetic induced dipoles whose polarizability is described below. We then introduce an optimization procedure to construct a nonzero spectral energy density that yields vanishing pairwise interactions, and test this procedure using the proposed model. Finally, we illustrate our approach with a realistic colloidal system, demonstrating that it is indeed possible to induce purely many-body interactions in practice. Interestingly, for a wide variety of optical responses, it possible to minimize the optically induced pair interaction potential to negligible values, well below the thermal energy $k_BT$, while the total electromagnetic energy density stays at arbitrarily high values, hence greatly enhancing the many body interactions over the pair ones.

 We consider a system comprising two identical absorptionless particles at positions $\textbf{r}_1$ and $\textbf{r}_2$, separated by a center-to-center distance $r=||\textbf{r}_2-\textbf{r}_1||$. Each particle is modeled as possessing both electric and magnetic dipole responses, characterized by scalar electric and magnetic polarizabilities $\alpha_e\left( \omega \right)$ and $\alpha_m\left( \omega \right)$, respectively. For each frequency $\omega$, the system is illuminated by an artificial random light field, constructed as a superposition of plane waves at $\omega$ with random wave vectors and polarization states distributed homogeneously and isotropically with an average squared amplitude given by $\left \langle|E_0|^2\right \rangle={2U_E}/({\epsilon_0\epsilon_h})$, where $U_E$ is the averaged electric energy density of the field. As established in \cite{setala_spatial_2003}, the cross-spectral density tensor of such a field is proportional to the imaginary part of the electromagnetic dyadic Green's tensor $G$ in the homogeneous host medium \cite{carminati_principles_2021} (See appendix A). Following the formalism introduced in \cite{brugger_controlling_2015} , for a given $u_E(\omega)$, the pair interaction induced by the random light field is conservative and given by the central potential 
 \begin{equation}
    U(r)=\int_0^\infty u_E\left(\omega\right)V\left(r,\omega\right)d\omega.
    \label{master-formula}
\end{equation}
Here, $V\left(r,\omega\right)$ is a function with units of volume accounting for the induced interaction at a single frequency that reads
\begin{equation}
        V\left(r,\omega\right) =\frac{2\pi}{k^3}\text{ImTr}\left[\mathbb{I}-k^4G\left(\textbf{r}_1,\textbf{r}_2,\omega\right)\alpha G\left(\textbf{r}_2,\textbf{r}_1,\omega\right)\alpha\right],
        \label{V-equation}
\end{equation}
where alpha is a $6\times6$ diagonal complex matrix defined as $\alpha\equiv\text{diag}\left(\alpha_e,\alpha_e,\alpha_e,\alpha_m,\alpha_m,\alpha_m\right)$.
Equation \eqref{master-formula} entirely describes the  pair potential contribution for each pair in an assembly of particles. Hence, if $U(r)=0$, at least within a range of distances, only many-body interactions remain in this range.

As a matter of principle, the energy density $u_E(\omega)$ could be selected at will and well above the Plank spectrum with the appropriate combinations of sources. Hence, from Eq. \eqref{master-formula} the resulting pair interaction potential $ U(r) $ can assume many different forms, in particular we could achieve a $ U(r)\ll K_B T$ even if the total energy density $U_E\equiv \int _0^\infty u_E\left(\omega\right)d\omega $ is arbitrarily large.

We model the electric and magnetic polarizabilities  $\alpha_d$, $d=e,m$ through a generic Lorentzian response

\begin{equation}
    \alpha_d\left(\omega\right) = \frac{6 \pi}{k^3} \frac{\gamma \omega}{\omega_0^2 - \omega^2 - i \omega \gamma}.
\label{lorentzian_pol}
\end{equation}

where $\omega_0$ is the resonance frequency, $\gamma$ is its damping rate, and $k$ the wave number. These polarizabilities satisfy the optical theorem for absorptionless particles \cite{jones2015optical} $k\,\mathrm{Im}\{\alpha_d\} = {k^4} |\alpha_d|^2/{6\pi}$.

To reduce the parameter space required to describe the polarizability, the frequency, wavenumber, and wavelength can be expressed in terms of their values at the electric resonance as
\begin{equation}
    \tilde{\omega}\equiv \frac{\omega}{\omega_0}, \quad \tilde{\lambda} \equiv \frac{\lambda}{\lambda_0} = \tilde{\omega}^{-1}.
\end{equation}
In addition, we define the resonance's quality factor as $ Q \equiv \frac{\omega_0}{\gamma}$, which let us rewrite the polarizability as
\begin{equation}
    \frac{\alpha_d(\tilde{\omega})}{\lambda_0^3} = \frac{3}{4 \pi^2  Q} \frac{\tilde{\omega}^{-2}}{1 - \tilde{\omega}^2 - i \tilde{\omega}/Q}.
\end{equation}
With this scaling, a single resonance is specified by its quality factor $Q$. 

\begin{figure}
    \centering
    \includegraphics[width=1\linewidth]{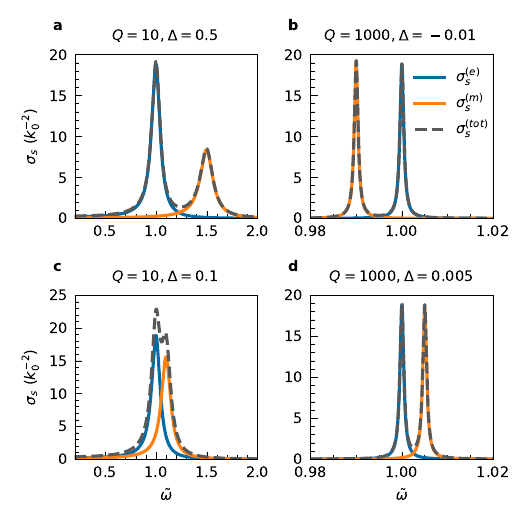}
    \caption{\textbf{a-d}: Scattering cross section $\sigma_s$ of particles modeled as electric and magnetic dipoles with Lorentzian polarizabilities, characterized by parameters $Q$ and $\Delta$. The contributions $\sigma_s^{(e)}$ and $\sigma_s^{(m)}$ are computed using only the electric ($\alpha_e$) or magnetic ($\alpha_m$) polarizability, respectively. $\sigma_s^{(\mathrm{tot})}=\sigma_s^{(e)}+\sigma_s^{(m)}$.}
    \label{fig:pol-examples}
\end{figure}

To describe both electric and magnetic polarizabilities independently, we introduce two quality factors, $Q_e$ and $Q_m$, along with a dimensionless detuning parameter $\Delta$ that sets their relative spectral position

\begin{equation}
    \Delta = \frac{\omega_0^m - \omega_0^e}{\omega_0^e},
\end{equation}
so that ${\omega_0^m}/{\omega_0^e} = 1 + \Delta$. In this work, except in the last part, we will set $Q=Q_e=Q_m$ to simplify the discussion. Figure \ref{fig:pol-examples} shows the electric and magnetic dipole contributions to the scattering cross sections, as well as the total one, obtained with the Lorentzian model for different combinations of parameters $Q$ and $\Delta$.

We represent the spectral energy density as a superposition of $N_l$ narrow lines at discrete frequencies $\tilde{\omega}_i$ and energy density contributions $U_E(\tilde{\omega}_i)$ :  
\begin{equation}
u_E(\tilde{\omega}) = \sum_{i=1}^{N_l} U_E(\tilde{\omega}_i)\,\delta(\tilde{\omega}-\tilde{\omega}_i).
\end{equation}
The induced pair potential as a function or the inter-particle distance reads
\begin{equation}
U(\tilde{r}) = \sum_{i=1}^{N_l} U_E(\tilde{\omega}_i)\,V(\tilde{r},\tilde{\omega}_i),
\end{equation}
where $\tilde{r}=r/\lambda_0$.  

For a given total electric energy density $U_E=\sum _i^{N_l}U_E(\tilde{\omega}_i)$, the relative contributions to the potential for each line is $C_i =U_E(\tilde{\omega}_i)/U_E$, so that we can rewrite the pair interaction as the energy density times  a volume  function $V_{\mathrm{tot}}(\tilde{r})$,
\begin{equation}
U_{\mathrm{tot}}(\tilde{r})=U_E V_{\mathrm{tot}}(\tilde{r}) = \sum_{i=1}^{N_l} C_i\,V(\tilde{r},\tilde{\omega}_i).
\end{equation}
For numerical optimization, we sample the potential at a discrete collection of distances $\tilde{r}_i$, $i=1,\cdots,N_r$, and we collect the sampled potential into a vector $\mathbf{V}_{\mathrm{tot}} = (V_{\mathrm{tot}}(\tilde{r}_1),\dots,V_{\mathrm{tot}}(\tilde{r}_{N_r}))^t$. 
We next define the loss function  $L = \lVert \mathbf{V}_{\mathrm{tot}} \rVert_2$ that should be minimized under the constrains of non-negative energy density for each line, $C_i \geq 0$, and constant total energy density of the field $\sum_i C_i = 1$. 
 In practice, this constrained optimization is carried out with the SLSQP algorithm in \textsc{SciPy}~\cite{2020SciPy-NMeth,kraft1988software}, 
using a tolerance of $10^{-9}$. We initialize the algorithm several times with different random guesses fulfilling the constrains in order to find the best minimum of the loss function.


We vary the quality factor $Q$ within the interval $Q \in [1,1000]$ and the detuning parameter $\Delta$ in an interval fulfilling $\Delta Q \in [0,5]$. $\Delta\ge 0$ since exchanging electric and magnetic polarizabilities does not alter the pair interaction (See appendix B).

For every pair of parameters $(Q,\Delta)$, we generate a set of $N_l = 200$ wavelengths $\tilde{\lambda}_i$ ($i=1,\dots,N_l$), evenly spaced within an interval $\tilde{\lambda}_i \in \left[\left({1+\Delta+2/Q}\right)^{-1},\,{\max(1-2/Q,0.3)}^{-1}\right]$. This construction ensures that the sampling remains well-defined even near resonance, while avoiding nonphysical values.
We consider an inter-particle distance range $\tilde{r}\in[0.5,3]$ discretized in $N_d=1000$ points to each pair of values $(Q, \Delta)$.

In order to assess the quality of the vanishing pair potential, we define a normalized loss function $\tilde{L}\equiv L/\|C_m V_m\|_2$, where $m$ is the index corresponding to the wavelength with the largest coefficient $C_m$, i.e. contributing the most to the total energy density.

\begin{figure}
    \centering
    \includegraphics[width=1\linewidth]{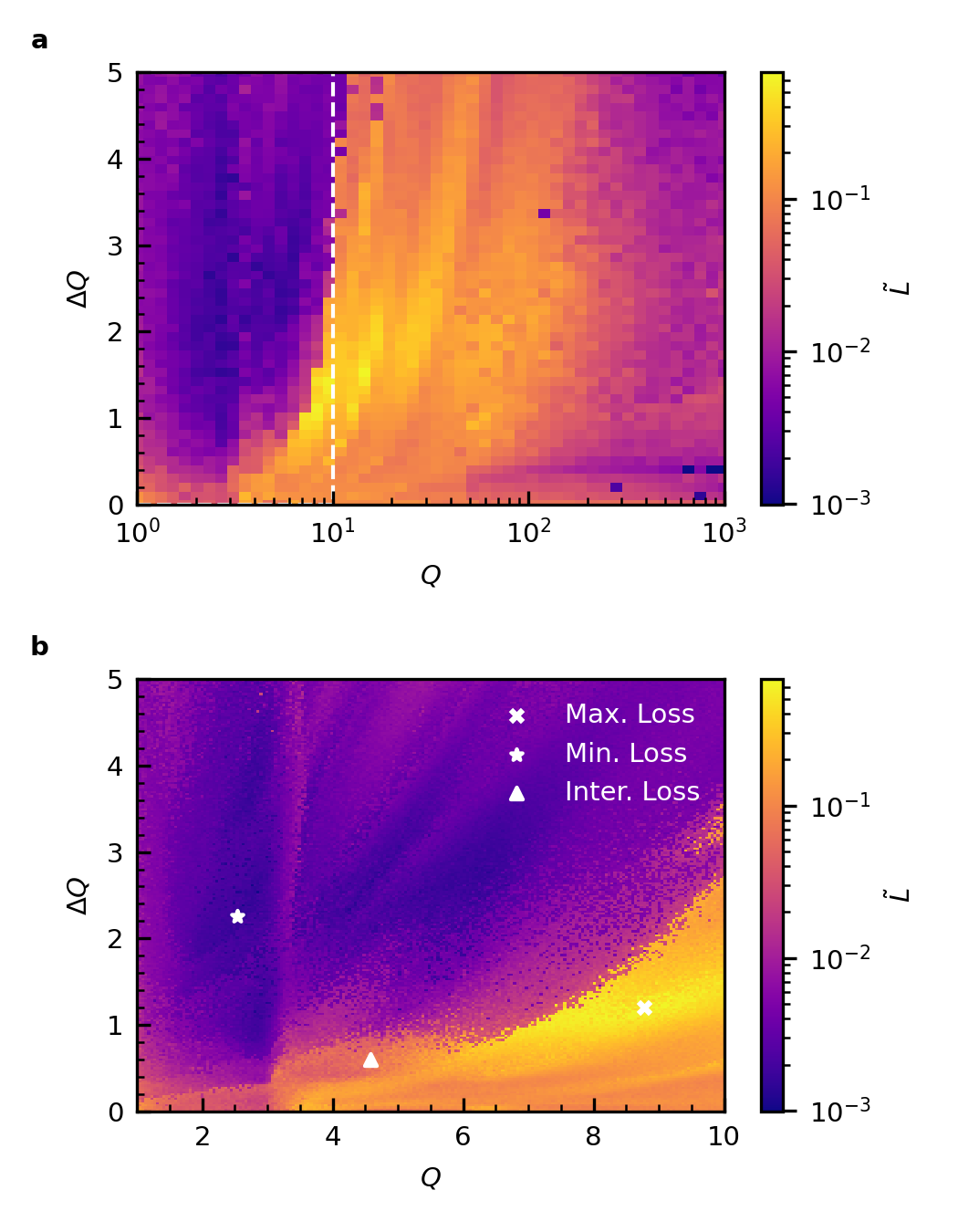}
    \caption{\textbf{a}: Normalized minimal loss function $\tilde{L}\equiv L/\|C_m V_m\|_2$ (see text) as a function of quality factor $Q$ of both electric and magnetic polarizabilities and detuning $\Delta$ between electric and magnetic resonances. \textbf{b}: Higher-resolution zoom on the region $Z \le 10$ indicated by the white dashed-square on panel \textbf{a}. The star and the  cross are showing the position of the minimum ($\tilde{L}=0.0098$) and maximum ($\tilde{L}=0.67924$) normalized loss functions, respectively. The triangle corresponds to a set of parameters giving an intermediate value of the normalized loss function ($\tilde{L}=0.06218$).}
    \label{fig:loss_maps}
\end{figure}

Figure \ref{fig:loss_maps}\textbf{a} shows the value of $\tilde{L}$ as a function of quality factor and detuning. On this map, we identify a region (when $Q\in[1,10]$ and $\Delta>0.4$) where the optimization yields very low residual pair interactions.

This region is shown in detail in Figure \ref{fig:loss_maps}\textbf{b}. We identify in this region the absolute minimum ($Q=2.54,\ \Delta Q=2.26$) and maximum $Q=8.78,\ \Delta Q=1.21$) of $\tilde{L}$ as well as an intermediate case.
 
This analysis shows that, although complete cancellation of pairwise interaction is generally not possible everywhere in the parameter space, wide and practically relevant regions exist where these interactions can be made negligible.

\begin{figure*}
    \centering
    \includegraphics[width=1\linewidth]{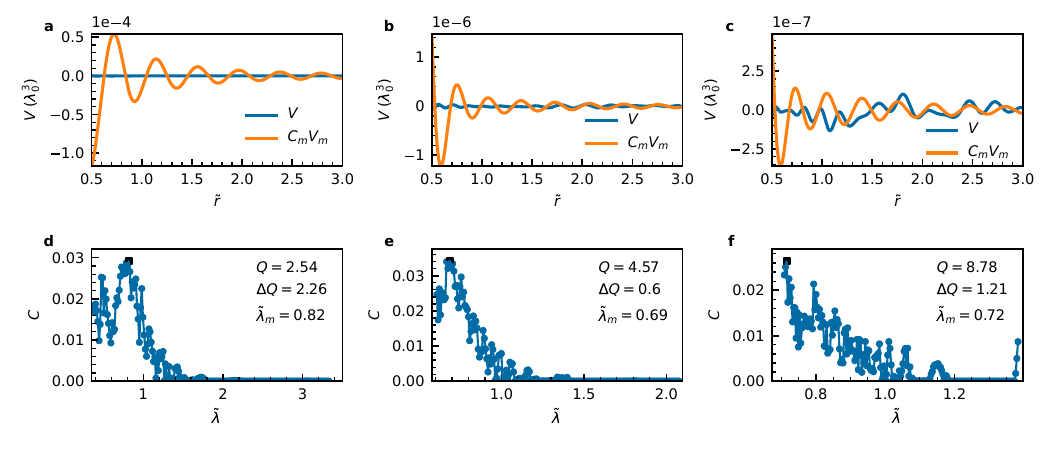}
    \caption{
Pair potential and energy density spectrum obtained at the point with the lowest found normalized cost function $\tilde{L}=0.0098$ (\textbf{a},\textbf{d}, resp.), at an intermediate value $\tilde{L}=0.06218$ (\textbf{b},\textbf{e}), and for the worst case $\tilde{L}=0.67924$ (\textbf{c},\textbf{f}) in figure~\ref{fig:loss_maps} . In all cases, the obtained potential $V(r)$ is compared with the contribution with maximum coefficient $C_mV_m$ marked as squares in the corresponding energy density spectrum. }    
    \label{fig:examples}
\end{figure*}

We examine in Figure~\ref{fig:examples} the three representative cases highlited in  Figure \ref{fig:loss_maps}\textbf{b}. 
The first case (Figure~\ref{fig:examples}\textbf{a},\textbf{d}) corresponds to the best performance found in parameter space. In this case, the potential $V$ appears essentially constant and close to zero, certainly at a much smaller value than the one corresponding to the maximum contribution $C_mV_m$. The second case (Figure \ref{fig:examples}\textbf{b},\textbf{e}) shows the potential obtained in a region of the parameter space where the loss function reaches intermediate values (triangle in fig.~\ref{fig:loss_maps}\textbf{c}). The third case (Figure~\ref{fig:examples}\textbf{c},\textbf{f}) shows the least successful optimization: Here the resulting potential reaches magnitudes comparable or even larger than those of the maximal coefficient contribution at large distances.

Despite the simplicity of the Lorentzian model, $Q$ and $\Delta$ can be chosen to closely reproduce the electric and magnetic polarizabilities of real homogeneous or core--shell spherical particles ~\cite{Paniagua2011Metallo}. To illustrate this point, we consider two dielectric particles with permittivity $\epsilon=12$, radius $a=230\,\mathrm{nm}$,  at a surface-to-surface separation $D$, and immersed in water ($\epsilon_h\simeq 1.77$).

For vacuum wavelengths in the range $1200$--$2200\,\mathrm{nm}$, the electromagnetic response can be accurately described by only their electric and magnetic polarizabilities \cite{garcia-etxarri_strong_2011, Fu2013Directional, Geffrin2012Magnetic}, $\alpha_e=i6\pi a_1 /k^3$ and $\alpha_m=i6\pi b_1 /k^3$, given by the first-order Mie coefficients \cite{bohren_absorption_2008,hulst_light_2012} , $a_1$ and $b_1$.

 We performed a least-squares optimization (using SciPy \cite{2020SciPy-NMeth}) of the parameters $Q_e$, $Q_m$, and $\Delta$, by fitting the scattering cross section spectrum from Lorenztian polarizabilities to the one obtained from Mie theory.

\begin{figure*}
\onecolumngrid
    \centering
    \includegraphics[width=1\linewidth]{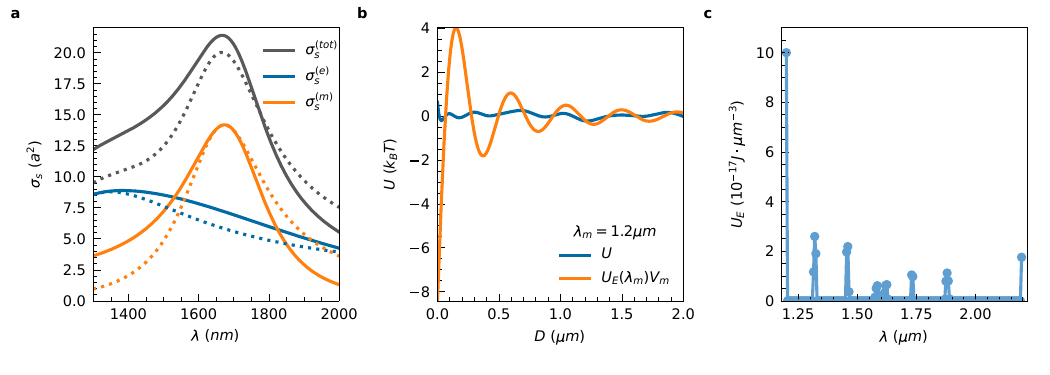}
    \caption{\textbf{a}: Comparison of the total, electric and magnetic contributions to the scattering cross section $\sigma_s$ (see legends) of a dielectric particle ($\epsilon=12$, radius $a=230nm$) computed with Mie theory (solid  lines) and with the best fit to a Lorentzian model (dotted lines). \textbf{b}: Minimized pair interaction $U\left ( D \right )$ (temperature $T=300\text{K}$) compared with the contribution with maximal energy density $C_m V_m$ as a function of the surface-to-surface distance $D$. \textbf{c}: Energy density spectrum required to minimize the pair interaction.}
    \label{fig:fig_si_part}
\end{figure*}

 The obtained result is shown in Fig.~\ref{fig:fig_si_part}\textbf{a}, where the optimal parameters $Q_e=2.2133$, $Q_m=5.7585$, and $\Delta=0.2302$ yield reasonable agreement. Notice that the electric resonance's wavelength is $\lambda_0=1384 nm$ and is determined using the maximum of $\sigma_s^{(e)}$ computed with Mie theory.

 With these parameters, we perform the minimization of the pair interactions. The minimized pair potential is presented in Fig.~\ref{fig:fig_si_part}\textbf{b}, and compared with the contribution with maximal energy density (highest point in the discrete spectrum in panel \textbf{c}). While the minimization may not be as good as in previous cases, this example shows that the residual pair interaction can still be forced to be well below $k_BT$.

In summary, we have proposed and tested a strategy to induce purely many-body interactions in colloidal systems using artificially generated random optical fields with an optimized spectral energy density that minimizes, and practically cancells in many cases, all pair interactions within a range of distances.

For this purpose it is essential to, at least, combine electric and magnetic dipole excitations in the electromagnetc response of the colloidal particles. This response has been modeled through a simple Lorentzian response that actually represents a good approximation in more realistic cases of Mie particles.

We have identified a wide range of parameters where the practical pair-interaction cancellation is feasible, allowing hence for the creation of driven systems whose dynamics is entirely dominated by pure many-body interactions. A detailed analysis of the dominant terms and scaling of the different many-body contributions will be presented in future work.

\section*{Acknowledgements}
We acknowledge useful and stimulating discussions with Frank Scheffold, Diego Romero-Abujetas and Manuel Marqués. Authors acknowledge the financial support from Schweizerischer Nationalfonds zur Förderung der Wissenschaftlichen Forschung (197146).

\bibliography{biblio.bib}
\section*{End matter}
\appendix 
\section{Electromagnetic Green's Tensor}
In free space, the dyadic Green tensors associated with electric and magnetic dipole radiation at position $\mathbf{r}_0$ are defined as
\begin{equation}
\begin{aligned}
G_E(\mathbf{r},\mathbf{r}_0) 
\equiv &\frac{e^{ikr}}{4\pi r} \Bigg[ 
\left( \frac{(kr)^2 + i kr - 1}{(kr)^2} \right)\mathbf{I} \\[4pt]
&+ \left( \frac{-(kr)^2 - 3ikr + 3}{(kr)^2} \right) 
   \hat{\mathbf{r}} \otimes \hat{\mathbf{r}} \Bigg].
\end{aligned}
\end{equation}

\begin{equation}
G_M(\mathbf{r},\mathbf{r}_0) \equiv \frac{e^{ikr}}{4\pi r}\, \left( \frac{ikr - 1}{kr} \right)\, \hat{\mathbf{r}}\times ,
\end{equation}
where $k$ is the wavenumber in the medium, $r = |\mathbf{r}-\mathbf{r}_0|$, and $\hat{\mathbf{r}} = (\mathbf{r}-\mathbf{r}_0)/r$. In this appendix, we omit the explicit dependence on $\omega$.
For compactness, these tensors can be gathered into the $6\times 6$ electromagnetic Green tensor
\begin{equation}
G(\mathbf{r},\mathbf{r}_0) = 
\begin{pmatrix}
G_E(\mathbf{r},\mathbf{r}_0) & i\,G_M(\mathbf{r},\mathbf{r}_0) \\[6pt]
-i\,G_M(\mathbf{r},\mathbf{r}_0) & G_E(\mathbf{r},\mathbf{r}_0)
\end{pmatrix},
\end{equation}
which is the one used in Eq. \eqref{master-formula}.

The cross-spectral energy tensor of the random field considered in this work can be written as \cite{setala_spatial_2003,brugger_controlling_2015}
\begin{equation}
\langle\mathbf{E}_0\left(\mathbf{r},\omega\right)\mathbf{E}_0^\dagger\left(\mathbf{r}',\omega'\right)\rangle=\delta\left(\omega-\omega'\right)\frac{8\pi U_E}{\epsilon_0\epsilon_hk}\text{Im}\left\{G_E\left(\mathbf{r},\mathbf{r}'\right)\right\},
\end{equation}
where $\epsilon_h$ is the dielectric constant of the medium and the electric energy density $U_E$ is a function of the amplitude $E_0$ of the plane waves generating the random field
\begin{equation}
    U_E=\frac{1}{2}\epsilon_0\epsilon_h|E_0|^2.
\end{equation}

\section{Expansion of $V$}
In the case where the two particles are aligned along $z$-axis, $V$, defined in E.q. \ref{V-equation} can be expanded as follows
\vspace{5pt}
\begin{widetext}
\begin{equation}
\begin{aligned}
V(r,\omega) = \frac{2\pi}{k^3} \Big\{ &
\ln\!\left( 1 - k^{4}\alpha_{e}^{2} G_{Ez}^{2}(r) \right) 
+ \ln\!\left( 1 - k^{4}\alpha_{m}^{2} G_{Ez}^{2}(r) \right) 
+ 2 \ln\!\Big[ \left( 1 - k^{4}\alpha_{e}\big(\alpha_{e} G_{Ex}^{2}(r) + \alpha_{m} G_{M}^{2}(r)\big) \right) \\[4pt]
&\times \left( 1 - k^{4}\alpha_{m}\big(\alpha_{m} G_{Ex}^{2}(r) + \alpha_{e} G_{M}^{2}(r)\big) \right) 
- k^{8}\alpha_{e}\alpha_{m}(\alpha_{e}-\alpha_{m})^{2} G_{Ex}^{2}(r) G_{M}^{2}(r) \Big] 
\Big\}.
\end{aligned}
\end{equation}
\end{widetext}

 $G_{Ex}\left(r\right)$, $G_{Ex}\left(r\right)$ and $G_{M}\left(r\right)$ are components of the electromagnetic dyadic Green's tensor that can respectively be written as
\begin{equation}
    G_{Ex}\left(r\right)=\left(1+\frac{i}{kr}-\frac{1}{k^2r^2}\right)\frac{e^{ikr}}{4\pi r},
\end{equation}
\begin{equation}
    G_{Ez}\left(r\right)=\left(\frac{-2i}{kr}+\frac{2}{k^2r^2}\right)\frac{e^{ikr}}{4\pi r},
\end{equation}
and
\begin{equation}
    G_{M}\left(r\right)=\left(i-\frac{1}{kr}\right)\frac{e^{ikr}}{4\pi r}.
\end{equation}

From these expression, it is clear that the interaction potential is invariant regarding the exchange of $\alpha_e$ and $\alpha_m$.


\end{document}